\documentclass{article}
\evensidemargin\oddsidemargin
\usepackage{latexsym}
\usepackage{amssymb,amsmath}
\usepackage{graphicx}
\newcommand{\x}{{\bf x}}

\newcommand{\IN}{{\mathcal  N}}

\newcommand{\IF}{{\mathcal  F}}
\newcommand{\R}{{\mathbf R}}

\newcommand{\Q}{{\mathbf Q}}
\newcommand{\ob}{\overrightarrow{\beta}}

\newcommand{\PP}{{\mathbf P}}

\begin{document}

\title{
{\bf Transcending the Limits of Turing Computability}\thanks{VAA  is
grateful for partial support
from the USA Civil Research and Development Foundation (CRDF grant
UM1-2090); CSC and BSP have been partially supported  by the
Auckland University Vice-Chancellor's  Development Fund 23124; BSP
was also partially  supported  by  JSPS(L02704) and RFFI(01149)
grants.}}
\author{Vadim A. Adamyan,\thanks{Department of Theoretical Physics, Odessa
National
University, ul. Dvoryanskaya 2, 65026 Odessa, Ukraine. E-mail: {\tt
vadamyan@paco.net.}} \ Cristian S. Calude\thanks{Department of Computer
Science, The
University of Auckland, Private Bag 92019, Auckland, New Zealand.
E-mail: {\tt cristian@cs.auckland.ac.nz}.} \ and Boris S.
Pavlov\thanks{Department of Mathematics, University of Auckland,
Private Bag 92019, Auckland, New Zealand. E-mail: {\tt
pavlov@math.auckland.ac.nz.}}}\date{\today}
\maketitle \thispagestyle{empty}
\begin{abstract}
Hypercomputation or super-Turing computation is a ``computation'' that
transcends the limit imposed by Turing's model of computability. The field
still faces some basic questions, technical (can we mathematically and/or
physically build a hypercomputer?), cognitive (can hypercomputers realize
the AI dream?), philosophical (is thinking more than computing?). The aim of
this paper is
to address the question:  can we mathematically  build a hypercomputer?
We will discuss the solutions of the Infinite
Merchant Problem, a decision problem equivalent to the
Halting Problem, based on results obtained in
  \cite{Coins,acp}. The accent will be on the new computational technique and
results rather than formal proofs.
        \end{abstract}

\section{Introduction}
Hypercomputation or super-Turing computation is a ``computation'' that
transcends the limit imposed by Turing's model of computability; for a
recent perspective see the special issue of the journal {\em Minds and
Machines}
(12, 4, 2002).
Currently there are
  various proposals to break Turing' Barrier  by showing that certain classes
  of {\it computing procedures}  have super-Turing power (see
\cite{casti97,karl,criscasti,cds,teuscher,copeland}). A specific
class of computing procedures,    \cite{etesi,Coins,kieu2,acp}
make  essential use of some physical theory,
relativity theory in \cite{etesi}, quantum theory
\cite{Coins,kieu2}; they  all reflect  an
attitude advocated by Landauer \cite{landauer-91,landauer99} ({\it
information is inevitably physical}) and Deutsch \cite{deutsch-85,
deutsch-97} ({\it the reason why we find it possible to construct,
say, electronic calculators, and indeed why we can perform mental
arithmetic \ldots is that the laws of physics ``happen" to permit
the existence of physical models for the operations of
arithmetic}).

The aim of the present paper is to revisit the solutions offered in
\cite{Coins,acp}.
We will focus on the novelty of the approach and we will discuss its power
and its limits. No proofs will be offered.

\section{The Classical  Merchant Problem}
  Recall that in the classical version
of the Merchant Problem we have $10$ stacks of  coins, each stack containing
$100$ coins, and  we know that {\it at most one stack contains
only false coins, weighting $1.01\,g$};   true coins
weight $1\,g$. The problem is to find the stack with false coins
(if any) {\it by only one weighting}. The classical solution
reduces the problem to  the weighting of a special
combination of coins: one coin from the first stack, two coins
  from the  second stack, \ldots , ten coins  from the tenth stack. If the
false
  coins are  present   in the  $N$-th stack, then
  the  weight of the  combination will  be  $55 + \frac{N}{100}\,g$;
otherwise
  the  weight is  just $55\,g$. The Merchant Problem  quoted
above was widely spread in allies armies during the Second World
War, cf. \cite{ignatov}. Probably the elegant
solution  described  above  was the very first solution of  a
computational problem bearing typical features of quantum
computing, see  an extended  discussion in \cite{Coins}.

  \section{The Infinite Merchant Problem}

In what follows we are going to consider the following generalization of the
problem,
  the Infinite Merchant Problem:
we assume that {\it  we have countable  many stacks, given in some
computable way, all   of
them, except at most one, containing   true  coins  only}. True coins
weight $1$ and false coins weight $1+2^{-j}$, $j>0$.
Again
we are allowed to take a coin from each stack and we want to
determine whether all coins are true or there is a stack of false
coins.

\if01
For computability and complexity theories see \cite{crisIR}, for quantum
computing see \cite{Gruska, wc,cp}. For scattering
theory see \cite{DG,MB};  \cite{meglicki} contains a detailed
discussion of  some applications of  scattering theory to quantum
computing.
\fi

Next we will show that the Infinite Merchant Problem is classically
undecidable by reducing it  to the Halting Problem, i.e. the problem to
decide whether an
arbitrary Turing machine (TM)
  halts on an arbitrary input. Assume that a TM operates on positive integers
and suppose, for the sake of contradiction, that there
exists  a TM  {\tt HALT} which can decide whether  a TM  $T$
(given by its code $ \#T$, a positive integer)
eventually stops  on  input $ x$:

\[{\tt HALT}(\#T, x)= \left\{ \begin{array}{ll}
1, & \mbox{if  $ T(x)$  stops,}\\
0, &  \mbox{otherwise.}
\end{array}
\right.\]
  We construct a TM  $Q$

  \[Q (x)= \left\{ \begin{array}{ll}
1, & \mbox{if  ${\tt HALT}(x,x) = 0, $  }\\
\mbox{loops forever}, &  \mbox{otherwise,}
\end{array}
\right.\]
and deduce the contradiction:
\[{\tt HALT}(\#Q, \#Q) = 1 \mbox{ iff } {\tt HALT}(\#Q, \#Q) = 0.\]

We next describe the reduction. Assume that we  have a classical solution of
the Infinite Merchant Problem and we are given
a TM $T$ and an input $x$ for $T$. We construct  a computable sequence $q_1,
q_2, \ldots q_i \ldots$ as follows: if the computation of $T(x)$ did not
stop till the $i$-th step, then we put
$q_i=1$; if the computation halted at step $i_0$, then we put $q_{i_0} = 1 +
2^{-j}$ and $q_r =1$,
for all $r > i_0$. The sequence $q_i$ satisfies all conditions of the
Infinite Merchant Problem and
$T(x)$   halts  if and only if there is a false coin, i.e.,  $q_{i_0} = 1 +
2^{-j}$, for some $j$. This shows that the Infinite Merchant Problem is
undecidable as the Halting Problem is undecidable.

In fact the two problems are equivalent. Indeed, assume that  we could
classically solve the Halting Problem.  To every sequence $(q_i)$ satisfying
the conditions of the Infinite Merchant Problem we  associate the  TM $T$
such that $T(i) = 1$ if $q_i=1$, and $T(i)=0$ otherwise. The TM $T'$ defined
by $T'(0) = \min \{i \mid T(i)=0\}$ halts at $0$ if and only if there an
$i_0$ such that $q_{i_0} = 1 + 2^{-j}$, i.e.,
$T'(0)$ halts if and only if there are false coins in the system. Hence,  a
classical solution of  the Halting Problem will produce a classical solution
for the  Infinite Merchant Problem.

The above discussion shows that undecidability is determined by the
impossibility to decide in a finite  time the answers to an infinite number
of questions, ``does the first stack contain a false coin?'',
``does the second stack contain a false coin?'', etc. This might be caused
either by the fact that the time of the computation grows indefinitely or
by the fact that the space of computation grows indefinitely or both. The
classical theories of computability and complexity (see, for example,
\cite{crisIR})  do not give any indication in this respect. In the following
section we will show that
time can be made finite provided we use a specific probabilistic strategy.

        \section{A  Probabilistic Solution}

      In this section we   present, in a slightly different way, the
      probabilistic solution  proposed in \cite{Coins}. We will adopt
        the following strategy. We are  given a
        probability $\theta = 2^{-n}$  and we assume that
      we work with a ``device" described below\footnote{As in
      \cite{Coins} we use
       quotation marks when referring to  our
       mathematical ``device".} with sensitivity given by a
        real $\varepsilon = 2^{-m}$.  Then, we
        compute classically a time $T=T_{\theta, \varepsilon}$
         and run the ``device" on a random input for the time $T$.
          If we get a click, then the system has false coins;
          if we don not get a click, then we conclude that with probability greater than
          $1-\theta$ all coins are true. An essential part of the method
          is the requirement that the time limit $T$ is {\em classically
computable.}

The ``device" (with sensitivity
$\varepsilon$) will distinguish the values of the iterated
quadratic form $\langle \Q^t (\x), \x\rangle = \sum_{i=1}^{\infty}
q_i^t  \, |x_i|^2,$
  by observing  the difference between averaging over trajectories
  of two discrete  random
  walks
with  two  non-perturbed  and perturbed sequences $t_l,\,
   \tilde{t}_l$ of  ``stops". The non-perturbed  sequence  corresponds
to  equal steps $\delta_m = 1$, $t_l = \sum_{m=0}^l \delta_m$, and
the  perturbed corresponds to the  varying  steps $\Delta_m,\, 0<
\Delta_m < \delta_m,\,\, \tilde{t}_l = \sum_{m=0}^l \Delta_m. $
   We  work with  the
   intersections of $l_2$  with the  discrete Sobolev class  $l_2^1$  of
square-summable sequences  with  the square norm
\begin{equation}
\label{norm1} \mid  \x \mid^2_1 \,\,\, = \, \sum_{_{_{m =
1}}}^{\infty} \,|x_{_m}- x_{_{m-1}} |^2,
\end{equation}
  and  the  discrete Sobolev class  $\tilde{l}_2^1$  of
weighted-summable sequences  with  the square norm
\begin{equation}
\label{norm2}  \parallel  \x
\parallel ^2_1 \,\,\,
= \, \sum_{_{_{m = 1}}}^{\infty}
\frac{1-{\Delta}_m}{{\Delta}_m}\,\,\,|x_{_m} - x_{_{m-1}}|^2.
\end{equation}

By natural extension from cylindrical sets we can define the
Wiener measures $\tilde{W}$ and $W$  on the spaces of trajectories
of  the  perturbed  and  non-perturbed random walks respectively
and use the   absolute continuity  $\tilde{W}$ with respect to
$W$: that is   for every $W$--measurable set $\Omega$,

\[
\tilde{W}(\Omega) = \frac{1}{\prod_{l=1}^{\infty} \sqrt{\Delta_l}}
\int_{\Omega} e^{- \sum_{m=1}^{\infty}
\frac{1-{\Delta}_m}{{\Delta}_m} \mid x_m - x_{m-1} \mid^2 } dW.
\]
Assume  that  the  ``device'' revealing the exponential growth of the
quadratic form  of  the  iterations  $\langle {\bf Q}^t (\x),\,\x \rangle$
clicks if
\[
\langle {\bf Q}^t (\x),\,\x \rangle \geq \, \parallel \x\parallel^2 +\,
\varepsilon \parallel \x\parallel^2_1.
\]

Thus the ``device'' sensitivity is defined in terms of the Sobolev norm.

Two cases may appear. If for some $T>0$, $\langle {\bf Q}^T (\x),\,\x
\rangle \geq \, \parallel \x\parallel^2 +\,
\varepsilon \parallel \x\parallel^2_1,$ then the ``device" has clicked and
we
know
for {\em sure} that there exist false coins in the system. However, it is
possible
that at some time $T>0$ the ``device"  hasn't (yet?) clicked because
$\langle {\bf Q}^t (\x),\,\x \rangle < \, \parallel \x\parallel^2 +
\varepsilon \parallel \x\parallel^2_1.$ This
may
happen because either all coins are true, i.e.,
$\langle {\bf Q}^t (\x),\,\x \rangle < \, \parallel \x\parallel^2 +
\varepsilon \parallel \x\parallel^2_1$, for all $t>0$, or because at time
$T$ the growth of $\langle {\bf Q}^{T}(\x), \x \rangle $ hasn't yet reached
the
threshold $\parallel \x\parallel^2 +
\varepsilon \parallel \x\parallel^2_1$. In the first case the
``device" will {\em never} click, so at each stage $t$ the test-vector $\x$
produces
``true" information; we can call $\x$ a  ``true"  vector. In the second
case, the
test-vector $\x$ is ``lying" at time
$T$ as we {\it do}  have false coins in the system, but they were not
detected
at time
$T$; we say that $\x$ produces ``false" information at time $T$.

If we assume that there exist false coins in the
  system, say at stack $j$, but the
   ``device" does not click at  the  moment  $T$, then the
   test-vector $\bf x$ belongs to the {\it indistinguishable set}

\[\IF_{\varepsilon, T} = \{ \x \in  l_2^1 \mid  ((1+\gamma)^T-1) \mid x_j
\mid^2 <
\, \varepsilon
\parallel\x \parallel^2_1, \, \mbox{for some   } j\}.
\]

In \cite{Coins} it was proven that  the Wiener measure of the indistinguishable set
tends to zero as $T\to \infty$:
\[\tilde{W}({\IF}_{\varepsilon,T})\le
\left(\frac{\varepsilon} { ((1+\gamma)^T -1-\varepsilon) \cdot
\prod_{m=1}^{\infty}\, {\Delta}_m }  \right)^{1/2}.\]

This fact is not enough to realize the scheme described in the beginning of
this section: we need a more  precise result,  namely we have to prove that
$\tilde{W}({\IF}_{\varepsilon,T})$  {\it
converges computably to zero.} And, indeed, this is true because:
$$
\tilde{W}({\IF}_{\varepsilon,T})\le \eta, \mbox{  provided   } t >
\log_{1+\gamma} \left(\frac{\varepsilon}{\eta^2 \,
\prod_{m=1}^{\infty}\, {\Delta}_m}+ 1+ \varepsilon\right).$$

Denote by $P(\IN)$ the  {\it a priori} probability of absence of
false coins  in the system. Then,  the {\it a
posteriori probability that the system contains only true coins,
when the ``device" did not click after running the experiment for the time $T$, is}
\[
{\mathbf P}_{\mbox{non-click}} (\IN)>  1 - \frac{1- P(\IN)}{P(\IN)} \cdot
\frac{\sqrt{\varepsilon}}{\sqrt{(1+\gamma)^T-1-\varepsilon} \,
  \sqrt{\prod_{m=1}^{\infty}\,{\Delta}_m}} \raisebox{.5ex}{.}\]

\section{A Brownian Solution Based on
Resonance Amplification}

In  \cite{Adamyan}    the  idea  to  consider a  single  act of
quantum  computation as a  scattering process was  suggested.

We  will first illustrate the method by describing
a simple   quantum  scattering  system  realizing the quantum
$C_{\mbox{NOT}}$  gate, i.e., a quantum gate
satisfying exactly to the same truth-table as the classical
controlled-NOT gate. The  $C_{\mbox{NOT}}$ device has two  input
and output channels. Each channel can be only in two different
states, say $\left| 0\right\rangle ,\left| 1\right\rangle $. The
\textit{in} and \textit{out} states of the control-channel are the
same, $\left| I_{in}\right\rangle =\left| I_{out}\right\rangle $,
but the \textit{in} and \textit{out} states of the current-channel may
be different, $\left| J_{in}\right\rangle \neq \left|
J_{out}\right\rangle, $  depending upon the state $\left|
J_{in}\right\rangle $ and the  control-channel state. The
classical controlled-NOT gate has the following truth-table:

\[
\begin{array}{cccc}
I_{in} & J_{in} & I_{out} & J_{out} \\
\hline\\[-2ex]
0 & 0 & 0 & 0 \\
0 & 1 & 0 & 1 \\
1 & 0 & 1 & 1 \\
1 & 1 & 1 & 0\\
\end{array}\]

  \noindent which describes the effect of the device on the above
\textit{in} states,

\[
\begin{array}{ccc}
\left| I_{in}\,\,J_{in}\right\rangle & \longrightarrow & \left|
I_{out}\,\,J_{out}\right\rangle \\
\hline\\[-2ex]
\left| 0\,\,0\right\rangle & \longrightarrow & \left| 0\,\,0\right\rangle \\
\left| 0\,\,1\right\rangle & \longrightarrow & \left| 0\,\,1\right\rangle \\
\left| 1\,\,0\right\rangle & \longrightarrow & \left| 1\,\,1\right\rangle \\
\left| 1\,\,1\right\rangle & \longrightarrow & \left| 1\,\,0\right\rangle
\end{array}
\]

  \if01
  (the first two qubits correspond
to {\it in} signals $I_{in}, J_{in}$ and the last two qubits
correspond to {\it out} signals $I_{out}, J_{out}$):

     \begin{center}
$C_{\mbox{NOT}}: \begin{array}{ccc}
|00 \rangle & \rightarrow &  |00 \rangle \\
|01 \rangle & \rightarrow &  |01 \rangle \\
|10 \rangle & \rightarrow &  |11 \rangle \\
|11 \rangle & \rightarrow &  |10 \rangle
\end{array}.$\end{center}
\fi

The quantum   $C_{\mbox{NOT}}$ gate operates not only on the
``classical"  states  $|0 \rangle $ and $|1\rangle $,
($C_{\mbox{NOT}} |ij\rangle = |ik\rangle $, where $i,j\in \{
0,1\}$ $k=i\oplus j$ (mod 2)),  but also     on all their  linear
combinations,
\[
  \alpha _{00}|00\rangle +\alpha _{01}|01\rangle +\alpha _{10}|10\rangle
+\alpha _{11}|11\rangle \longrightarrow \alpha _{00}|00\rangle +\alpha
_{01}|01\rangle +\alpha _{10}|11\rangle +\alpha _{11}|10\rangle.
\]
This quantum transformation  can be  presented  via the  unitary matrix

\begin{equation}\label{unitarymatrix}
         U=     \left( \begin{array}{cccc}
      1 & 0 &  0 &  0 \\
0 & 1 &  0 &  0 \\
0 & 0 &  0 &  1 \\
0 & 0 &  1 &  0
\end{array} \right),
\end{equation}

\noindent    with  respect  to  the canonical  basis $ (e_1, e_2, e_3, e_4)
= (|00\rangle,\, |01\rangle,\, |10\rangle,\, |11\rangle)$. More
importantly, $C_{\mbox{NOT}}$  in combination with all 1-qubit gates is  {\it
universal }\footnote{Every classically computable function can be
computed by a small universal set of gates like $\{$OR,  NOT$\}$
or $\{$NAND$\}$.  A set of quantum  gates ${\mathcal  S}$ is called
universal  if any unitary operation can be approximated with an
arbitrary accuracy by a quantum circuit involving gates in ${\cal
S}$; see more in \cite{feynman85, feynman96,Gruska,cp}.} and
{\it it cannot be written as a tensor product of  two binary
operators  $U = U_1 \, U_2,\,\, U_l \neq I$.}

We  claim that the  matrix $U$ in (\ref{unitarymatrix}) can   be
realized as  a  {\it scattering  matrix} of  a special  {\it
quantum dot}.  First, here is the motivation. Consider Figure~1 in
which two isolated quantum wires are placed in proximity and there
is   a window region in which the two wires are coupled. An
electron moving in the window region oscillates between the two
quantum wires and the probability of the electron exiting into a
specific quantum wire depends on the length of the window. This
  ``switching  phenomenon" was  discovered by del
Alamo and Eugster   \cite{Alamo} and  intensely discussed
in literature, see for  instance \cite{Exner,Antoniou}.
  We can arrange the setup in such a way that, under normal
conditions,
  the electron exits from the same wire it enters, but switches to the
   other wire when a classical extra potential is applied, a realization of
   the $C_{\mbox{NOT}}$ relay. If the control is quantum too, then we obtain
the $C_{\mbox{NOT}}$ gate. \\

\begin{figure}[h]
\begin{center}
\includegraphics[width=2in]{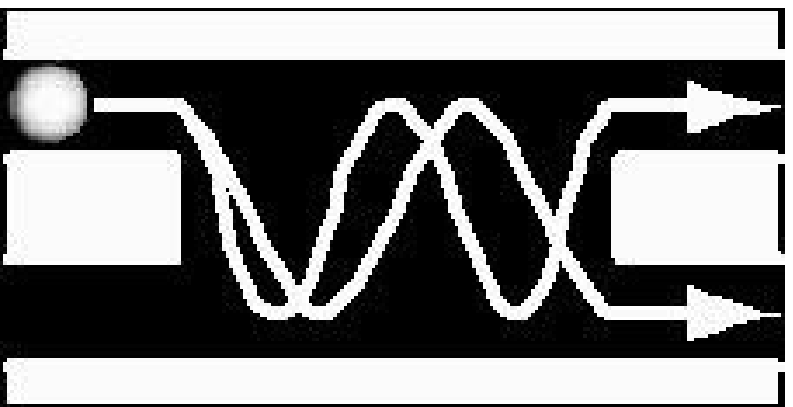}
\caption{$C_{\mbox{NOT}}$ as quantum dot}
\end{center}
\end{figure}

We continue with the mathematical model and  assume  that the
quantum dot is inserted in an one-dimensional quantum wire  ${\bf
R }= (-\infty,\, \infty)=  {\bf
R}_+ \, \cup \,{\mathbf R}_- $  between the wires  ${\mathbf R}_+
\,,\,{\mathbf R}_-$  and   a single electron may be redirected
according to the state of the dot.

We assume also that the {\it
inner} Hamiltonian of the quantum dot is presented by a finite
or an infinite diagonal matrix $ A =\mbox{diag}\,\left\{
 \alpha_1^2, \alpha_2^2,\dots ,\alpha_N^2, \ldots \right\}$
 with  positive  diagonal  elements $\alpha_1^2
<\alpha_2^2 < \alpha_3^2< \dots <\alpha_N^2 < \ldots$.
\if01
We assume also that the {\it
inner} Hamiltonian of the quantum dot is presented by a finite
diagonal matrix which is either   $ A_1 =\mbox{diag}\,\left\{
\alpha_1^2, \alpha_3^2,\dots ,\alpha_N^2 \right\}$ or $A_2
=\mbox{diag}\,\left\{\alpha_2^2,\alpha_3^2,\dots,\alpha_N^2
\right\}$, with  positive  diagonal  elements, $\alpha_1^2
<\alpha_2^2 < \alpha_3^2< \dots <\alpha_N^2$. 
\fi
We  assume  that the
quantum  dot  is inserted  in  an one-dimensional quantum  wire
$-\infty< x< \infty$ at the origin  and a  proper  boundary
condition is satisfied  (see (\ref{one}))  for
connecting it  with the  Schr\"{o}dinger operator on  the  wire
defined  in  the  space  of square-integrable vector-functions
$L_2 (\R_+ , {\mathcal E})$ with values in the infinitely-dimensional Hilbert space
${\mathcal E}$
\begin{equation}
\label{S_wire}
l = -\frac{d^2}{dx^2} \cdot
\end{equation}
One could  assign to the  above quantum  system a product
space  $H_e\otimes H_d$ constituted respectively by the states  of
the electron and  the  states of the dot,  and consider an
evolution of the system generated  by the total Hamiltonian ${\cal
H}_e + {\mathcal  H}_d + {\mathcal  H}_{int}$ with a proper interaction
term.
This would  lead to a  quite sophisticated problem of quantum
mechanics, similar to three-body problems, see for instance
\cite{MP95}. We  assume now that  {\it the  state  of the dot is
selected independently} and thus reduce the above problem to the
corresponding {\it one-body} problem for an electron scattered in
the quantum wire depending on the state of the dot. The
corresponding device should be called rather {\it quantum relay} rather
than  {\it quantum gate}; however,  it  may be  transformed into a
quantum gate if the state of  the dot is obtained as  a quantum
state with  finite life-time. Practically  the model suggested
below is acceptable if the  life-time of  the state of the dot is
long enough during the scattering experiment.The
corresponding general  ``zero-range"  quantum Hamiltonian
(solvable  model)  is described   as a
self-adjoint extension ${\bf A}_{\beta}$ of the  orthogonal sum
$l \oplus A$ restricted to $ l_0\oplus A_0$ in $L_2(\R, {\mathcal E})
\oplus E $  onto a proper domain; here  ${\mathcal  E}$ is the infinitely-dimensional input space and $E$ is the inner space (with  dim $(E)\ge 2$). 
The  {\it spectrum} $\sigma_{\beta}$ of  the operator ${\bf
A}_{\beta}$ is absolutely-continuous and  fills the positive
half-axis $\lambda \geq 0$ with multiplicity  dim $({\mathcal  E})$.
The  role of eigen-functions of the spectral point  $p^2 = \lambda
>0$ is played by the {\it scattered waves}
$\overrightarrow{\Psi}_{\nu},\, \overleftarrow{\Psi}_{\nu}$,
labeled  with  vectors $\nu \in {\mathcal  E}$. The  components of the
scattered waves $\Psi_{\nu} (p)$ in  the  {\it outer space} $L_2
(\R)$ are presented  as  linear  combinations of exponentials:
\[
\overrightarrow{\Psi}_{\nu,p} (x) = \left\{
\begin{array} {cc}
e^{-ipx} \nu + e^{+ipx} \overleftarrow{R}(p) \nu,\,\, & x<0,\\
e^{-ipx} \overrightarrow{T}(p) \nu,\,\, & x>0,
\end{array}
\right.
\]
\begin{equation}
\label{S-waves} \overleftarrow{\Psi}_{\nu,p} (x) = \left\{
\begin{array} {cc}
e^{ipx} \nu + e^{-ipx} \overrightarrow{R}(p) \nu,\,\, & x>0,\\
e^{ipx} \overleftarrow{T}(p) \nu,\,\, & x<0.
\end{array}
\right.
\end{equation}
The matrix
        \begin{equation}
        \label{S_matr} {\bf S}_{\beta}(p) = \left(
        \begin{array}{cc}
        \overrightarrow{T}(p) & \overrightarrow{R}(p)\\
        \overleftarrow{R}(p) & \overleftarrow{T}(p)
         \end{array}
        \right),
        \end{equation}
is  called the {\it scattering matrix} of the operator ${\bf
A}_{\beta}$.\footnote{The  transmission coefficients appear on the
main  diagonal of  the  matrix to fit the physical meaning of the
scattering  matrix for small values of $|\beta|$, when it is
reduced  to the undisturbed transmission ${\bf S}(p) =  I$.}

The evolution of the {\it wave function} of  the  quantum
mechanical system  with    Hamiltonian ${\bf A}_{\beta}$ given by
the equation
   \begin{equation}
\label{evolut}
\frac{1}{i}\frac{\partial \Psi}{\partial t} =  {\bf A}_{\beta} \Psi,
\end{equation}
and  proper  initial  condition
\[
\Psi \bigg|_{t=0} = \Psi_0,
\]
can be described by the correspopnding evolution operator
constructed from the above scattered waves and square-integrable
bound states $\Psi_{s}$ which satisfy the homogeneous  equation
\[
{\bf A}_{\beta} \Psi_{s} = \lambda_{s} \Psi_{s},
\]
with  negative  eigen-values  $\lambda_s$. Bound  states  do  not
play an essential role in   our construction,  so we may assume
that  the  initial  state $\Psi_0$ is orthogonal to all bound
states  and  may  be expanded in an analog  of Fourier integral
over  the  scattered waves
\[
\Psi_0 =  \frac{1}{2 \pi} \int_R \sum_{\nu}  {\Psi}_{\nu,p}\langle
{\Psi}_{\nu,p}, \Psi_0\rangle dp.
\]
Then the evolution described  by the  solution of the equation
(\ref{evolut})
and the above initial data  can  be presented  as a (continuous) linear
combination
\[
\Psi (t) =  \frac{1}{2 \pi} \int_{\R} \sum_{\nu} e^{ip^2 t}
{\Psi}_{\nu,p}\langle {\Psi}_{\nu,p}, \Psi_0\rangle dp
\]
of  {\it modes} incoming from infinity on the
        left $(-\infty)$ and on the right $(+\infty)$,  and
        outgoing modes
{\it scattered} to  both  directions  $ \pm\infty $
       according to the solution  of the time-dependent Schr\"{o}dinger
        equation $\frac{1}{i}\frac{\partial {\bf U }}{\partial t } = {\bf
        A}{\bf U}:$
        \begin{equation}
        \label{time_dep}
   {{\bf S}_{\beta}(p)}:
         e^{ip^2 t}
         \left(
        \begin{array}{c}
        e^{-ipx} \nu_{left}\\ e^{ipx} \nu_{right}
        \end{array}
        \right)
         \longrightarrow
        e^{ip^2 t}
         \left(
         \begin{array}{cc}
         e^{-ipx} (\overrightarrow{T}(p)\nu_{left}\phantom{+}  +&
        \overrightarrow{R}(p)\nu_{right})\\
          e^{ipx}  (\overleftarrow{R}(p)\nu_{left} \phantom{+} + &
        \overleftarrow{T}(p)\nu_{right})
         \end{array}
          \right).
        \end{equation}

The analytic structure  of  the  scattering  matrix  depends
upon the  structure  of  the  inner  Hamiltonian and  a  sort  of
interaction   between  the   inner  $E$  and  outer
$L_2 ({\mathbf R}, {\mathcal E}) $ spaces. To avoid the  discussion of the general
situation  we use here a  scattering  matrix for a solvable model of 
the quantum
dot  which  is  based  on  zero-range potential  with inner structure,
introduced in \cite{Zero-range} and  already  used  in  \cite{Antoniou}
for the description of typical  features  of  nano-devices, see
  also   \cite{acp}.

   If  we  choose  an infinitely-dimensional  input space
${\mathcal  E}\oplus {\mathcal  E}$ with components corresponding  to  the  amplitudes  of
  the  scattered  waves  at  $\pm \infty$
  and an one-dimensional {\it deficiency
        subspace}    ${\mathcal  N}_i$  spanned
        by  the normalized  vector $e = e_i \in {\mathcal E}$ (see \cite{acp})
        and  introduce  the  scalar function
      $${\mathcal  M} = \langle \frac{I + \lambda A}{A - \lambda I}e,e
        \rangle,$$ then using the interaction defined by the
        boundary  conditions (\ref{one}) depending on a vector $\beta \in {\mathcal E}$
imposed  on the  boundary  values (the  jump $[u'](0)$ and  the value $u(0)$
at  the  origin) of the  component  of the  wave-function in the 
outer  space
and the {\it symplectic coordinates}  $\vec{\xi}_{\pm} \in {\mathcal E}$ of the
inner  component  of  the  wave-function (see \cite{boris}):

       \begin{equation}
        \label{one}
         \left(
        \begin{array}{c}
        [u'](0)\\ -\vec{\xi}_{-}
        \end{array}
        \right)
        =
         \left(
        \begin{array}{cc}
        0 & \beta \\ \beta^+ & 0
        \end{array}
        \right)
         \left(
        \begin{array}{c}
        u(0) \\ \vec{\xi}_{+}
        \end{array}
        \right),
        \end{equation}

    \noindent     we  obtain the scattering matrix in the form
        \[
         {\bf S}_{\beta}(p) = \left(
        \begin{array}{cc}
        \overrightarrow{T}(p) & \overrightarrow{R}(p)\\
        \overleftarrow{R}(p) & \overleftarrow{T}(p)
         \end{array}
        \right),
       \]
       with  equal  transmission  and  reflection coefficients
$\overrightarrow{T},
       \overleftarrow{T}, \overrightarrow{R}, \overleftarrow{R}$:

       \[
        \overrightarrow{T}(p) =  \overleftarrow{T}(p) = \PP^{\bot}_{\beta} + 
        \frac{2ip}{2ip + |\beta|^2 {\mathcal  M}^{-1}} \PP_{\beta}, 
       \,
        \overrightarrow{R}(p) =  \overleftarrow{R}(p)=-
        \frac{|\beta|^2 {\mathcal  M}^{-1}}{2ip  +  |\beta|^2 {\cal
        M}^{-1}} \PP_{\beta}. 
        \]
        
Here $\PP_{\beta}$ is the orthogonal projection of ${\mathcal E}$ onto the one-dimensional space spanned by the vector $\beta$, and  $\PP^{\bot}_{\beta} = I -\PP_{\beta} $ is the orthogonal projection on the complimentary space.
The constructed  solvable  model  reveals the  role of
        zeroes  of  the  scattering  matrix  -- the
        {\it resonances} --
        in  implementing the  switching function.

       \if01
 Next  we  explore the properties of the  scattering matrix
        depending  on  distribution  of  occupied  and  non-occupied
        levels  in  the  ``quantum dot"  described  by  the  inner  Hamiltonian.
   We  observe  first the behaviour of  the scattering  matrix at
   the  resonance energy
        $\alpha_1^2 >0$ in  case    the resonance level
        $\alpha_1^2$ in the  quantum dot is  vacant,  but $\alpha_2^2$
        is occupied, hence eliminated  from the  quantum  picture
         (due to Pauli's exclusion principle, as discussed below).
        In this  case we  have
         $${\mathcal  M}_1= \frac{1+ \alpha_1^2 \lambda}{\alpha_1^2 -\lambda}
        |e_1|^2  + \sum_{l = 3}^N \frac{1+ \alpha_l^2
\lambda}{\alpha_l^2 -\lambda}
        |e_l|^2 =
         \frac{1+ \alpha_1^2 \lambda}{\alpha_1^2 -\lambda}
        |e_1|^2  +  {\mathcal  M}_3,
        $$ where $|e_l|^2$ are the squares of  the  Fourier coefficients of
        the  deficiency  vector $e$ with respect to the  eigen-vectors
        of  the  operator  $A$.

        Next we consider the case when the resonance level $\alpha_1^2$
        is occupied, but the level $\alpha_2^2$ is vacant. In  this
case  $${\mathcal  M}_2 =
        \frac{1+ \alpha_2^2 \lambda}{\alpha_2^2 -\lambda} |e_2|^2
        + \sum_{l = 3}^N \frac{1+ \alpha_l^2 \lambda}{\alpha_l^2 -\lambda}
        |e_l|^2 =       \frac{1+ \alpha_2^2 \lambda}{\alpha_2^2 -\lambda}
|e_2|^2 +
        {\mathcal  M}_3.
        $$
      where $|e_l|^2$ are the squares of the Fourier coefficients
       of  the  deficiency  vector  with
        respect to the  eigen-vectors of  the  operator $A_2$.

 The above constructed  model  corresponds  to  ``spin-less"
        electrons (electrons  with the constant spin  in absence of
        the magnetic field) 
\fi

We  observe  first the behaviour of  the scattering  matrix at
   the  resonance energy
        $\alpha_1^2 >0$ in  case    the resonance level
        $\alpha_1^2$ in the  quantum dot is  vacant as well as all levels
        $\alpha_l^2$ above $\alpha_1^2$.
        In this  case we  have
         $${\mathcal  M}_1= \frac{1+ \alpha_1^2 \lambda}{\alpha_1^2 -\lambda}
        |e_1|^2  + \sum_{l = 2}^N \frac{1+ \alpha_l^2
\lambda}{\alpha_l^2 -\lambda}
        |e_l|^2 =
         \frac{1+ \alpha_1^2 \lambda}{\alpha_1^2 -\lambda}
        |e_1|^2  +  {\mathcal  M}_3,
        $$ where $|e_l|^2$ are the squares of  the  Fourier coefficients of
        the  deficiency  vector $e$ with respect to the  eigen-vectors
        of  the  operator  $A$.

                   Next we consider the case when the resonance level $\alpha_1^2$
        is occupied. In  this
case  $${\mathcal  M}_2 =
        \frac{1+ \alpha_2^2 \lambda}{\alpha_2^2 -\lambda} |e_2|^2
        + \sum_{l = 3}^N \frac{1+ \alpha_l^2 \lambda}{\alpha_l^2 -\lambda}
        |e_l|^2 =       \frac{1+ \alpha_2^2 \lambda}{\alpha_2^2 -\lambda}
|e_2|^2 +
        {\mathcal  M}_3.
        $$
     
 In the above analysis we have ignored the electron spin, that is we have assumed that  all  electrons  have the same constant spin
 on the quantum circuit
        ${\mathbf R}_- \cup {\mathbf R}_+$,
        with the quantum dot attached.\footnote{Our hypothesis is satisfied in case the travelling electrons
  are polarized and  the electron on the level
 $\alpha_{1}^2$ is polarized. Note that Pauli's  exclusion principle
is still valid, but  with only  one electron on each  orbital:
the magnetic field is absent, so the polarization is not changed during
the experiment.}  We  may  assume that  the  circuit
        lies on the surface
        of a semiconductor with  Fermi-level  $\alpha_1^2$ (see
        \cite{Madelung}).
 
 \if01
 ;  the  levels
         $\alpha_1^2,\,\,\alpha_2^2$  are (due to Pauli's exclusion  principle)
                alternatively  occupied  by  electrons travelling
        on the circuit or by  electrons transferred from one level
        to  another inside the  quantum dot under the resonance laser
        shining.\footnote{The  manipulation
        of  the  resonance  quantum
        dot is  a  sophisticated few-body
        problem of  quantum  scattering. A  ``solvable" model for
        it was  discussed  in  \cite{MP95}. The model involving
        the  resonance  laser  shining  as  a  tool of manipulation
        of  the  current through  the quantum
        dot was introduced  in
        \cite{Antoniou}.}
\fi
 
 We assume that the state of the dot  with  the  level  $\alpha_1^2 $ vacant  corresponds
        to  $I_{in}= I_{out} = 0$ and  the  state of the dot with the
        level $\alpha_1^2 $  occupied corresponds  to  $I_{in}= I_{out} = 1$. We  identify
   these  states  of  the  system as    state  $S_{1}$ and  state
$S_{2}$, respectively.

        For  every  vector  $\beta$   the  transmission  coefficients
         on  the  resonance electron's energy $\lambda=\alpha_1^2$
        can be expressed as (see \cite{acp}):
         \[
           \overrightarrow{T}(p) =  \overleftarrow{T}(p) = \PP^{\bot}_{\beta} + 
        \frac{2ip}{2ip  + |\beta|^2 {\mathcal  M}_1^{-1}} \PP_{\beta} = I.\,\,
        \overrightarrow{R}(p) =  \overleftarrow{R}(p)=0,
        \]
        \noindent
   (at the  resonance  energy we  have ${\mathcal  M}_1^{-1} = 0$), so the scattering
 matrix becomes the identity when $\alpha^{2}_{1}$ {\it is not occupied}.

    In  the  second case, when  the resonance
        level  $p^2 = \lambda = \alpha_1^2$ is  occupied, we  obtain
        (due again to  Pauli's  exclusion principle)  the
        following  expression for  the  transmission  coefficients
        of  passing  electrons  with  resonance energy:
       \[
       \overrightarrow{T}(p) =  \overleftarrow{T}(p) =
        \frac{2ip {\mathcal  M}_2(\lambda)}{2ip{\mathcal  M}_2(\lambda)  -
|\beta|^2} =
         \frac{2ip \left(
        \frac{1+ \alpha_2^2 \alpha_1^2}{\alpha_2^2 -\alpha_1^2} |e_2|^2
        + \sum_{l = 3}^N \frac{1+ \alpha_l^2 \alpha_1^2}{\alpha_l^2
-\alpha_1^2}
        |e_l|^2 \right)}{2ip \left(
        \frac{1+ \alpha_2^2 \alpha_1^2}{\alpha_2^2 - \alpha_1^2} |e_2|^2
        + \sum_{l = 3}^N \frac{1+ \alpha_l^2 \alpha_1^2}{\alpha_l^2
-\alpha_1^2a}
        |e_l|^2 \right) + |\beta|^2 } \raisebox{.5ex}{,}
        \]
        and the corresponding  expressions  for the  reflection
        coefficients
        \[
         \overrightarrow{R}(p) =  \overleftarrow{R}(p)= - \frac{|\beta|^2}
         {2ip {\mathcal  M}_2 + |\beta|^2} \PP_{\beta}, 
         \]
        which can be approximated, for  large enough $\beta$,  as
        \[
        \overrightarrow {T}(\alpha_1) =  \overleftarrow{T}(\alpha_1)
\approx  \PP^{\bot}_{\beta},\,\,
         \overrightarrow {R}(\alpha_1) =  \overleftarrow{R}(\alpha_1) \approx
         -\PP_{\beta}.
        \]
        Hence the  scattering matrix is equal to
           \begin{equation}
           \label{resonance1}
            {\bf S}_{\beta}(\alpha_1^2) = \left(
        \begin{array}{cc}
        \PP^{\bot}_{\beta} & - \PP_{\beta}\\
        -\PP_{\beta} & \PP^{\bot}_{\beta}
         \end{array}
        \right) = I - 
 \left(
        \begin{array}{cc}
        \PP_{\beta} & \PP_{\beta}\\
        \PP_{\beta} & \PP_{\beta}
         \end{array}
        \right),
   \end{equation}
        for   relatively large enough  $\beta$, if  the  resonance
        level $\alpha^2_1$ {\it is  occupied}.

We continue by showing how the probabilistic approach discussed in the
previous
section can be realised.
\if01
We consider now
an imaginable quantum scattering system with an
infinitely-dimensional input-space for which  a solvable model was
described
above
as an
extension of the orthogonal sum $A_0 \oplus l_0 $ with boundary
condition (\ref{one}). 
\fi
We consider now an imaginable quantum scattering system with an
infinitely-dimensional input-space and, in particular, with the infinite dimensional space  $\mathcal E$
for which  a solvable model was described above as an extension of the orthogonal sum $l_0 \oplus A_0 $ with
boundary condition (\ref{one}), in which , however,  the $\beta$-channel connecting the  outer subspace
$ L_2(\R, {\mathcal E})$ with the inner subspace $E$ is as before two-dimensional.
We associate  these extensions with two
states $(S_{1}, S_{2})$ of the total quantum system combined of  the
inner and outer components, with  the interaction respectively
{\it switched  on} via the boundary condition (\ref{one}),
$\beta\neq 0$, or  {\it switched off}, $\beta = 0 $, and we interpret the
Halting Problem in a probabilistic  setting   as  the problem  of
distinguishing  of  the  states
    $(S_{1}, S_{2})$
    of  the  quantum  system  via  a  scattering  experiment  with
    a  random  input.

\if01
   In  scattering  experiments  one  usually observe  the  asymptotic
    behaviour of    scattered  waves - the  eigen-functions  of
    the  absolutely-continuous  spectrum  of  the  operator 
    at  infinity. This  behaviour  is  defined  in  our case  by
     the  corresponding  reflection coefficients
      (``stationary" scattering matrices), which  are
    different   depending upon   the ``state" of  the  system.
     The reflection coefficient in the first state
   can  be calculated via separation of variables. The calculation of the reflection coefficient in the second
case is reduced to matching the boundary data  $\vec{\xi}_{\pm}$ of the solution of the
adjoint equation $A_0^+ u = \lambda u$ in the inner  space $E$
with the boundary data of the Anzatz
combination  of incoming and  outgoing  modes.

In a previous example   we have considered the
  scattering  matrix  with only  two eigen-values $\pm 1$.
  Applying the  same  straightforward  approach
  based  on  direct  comparison  of  the  correlation
  $\langle {\bf S}{\bf x},\,{\bf x}\rangle$ with
  $\langle {\bf x},\,{\bf x}\rangle$  to the  general  case
when the  eigen-values  $e^{is_l}$ of  the  scattering  matrix  are
complex, gives  the  quadratic form
\[
\langle {\bf S}{\bf x},\,{\bf x}\rangle- \langle {\bf x},\,{\bf x}\rangle
=\sum_l (e^{is_l} - 1)|x_l|^2,
\]
which  cannot  be easily estimated.  An even more  difficult
problem appears  if we try  to  estimate the  correlation of
the  iterated  scattering  matrix
\[
\langle {\bf S}^m{\bf x},\,{\bf x} \rangle - \langle {\bf x},\,{\bf
x}\rangle =\sum_l (e^{i m s_{l}} - 1)|x_l|^2,\, m=2,3,\dots.
\]

In a different approach we consider  the  above  solvable  model with rational function  ${\mathcal M}
(\lambda)$. It  has  a  positive imaginary  part in  upper  half-plane  $\Im
\Lambda >0 $ and  negative  imaginary part in lower  half-plane
$\Re\lambda <0$. Then,  the two by two  scattering  matrix  can  be  presented
as  a  funnction of  momentum  $\zeta = \sqrt{\lambda}$
via  orthogonal projections ${\mathbf P}_{sym},\, {\mathbf P}_{asym}$ in
the  input space ${\mathcal E}$ onto  the  subspace  of  symmetric and
asymmetric  input  vectors
\[
{\mathcal E}_{sym} = 
\left(
\begin{array}{c}
1\\1
\end{array} \right),\,\,
{\mathcal E}_{asym} = 
\left(
\begin{array}{c}
1\\-1
\end{array} \right),
\]
\begin{equation}
\label{fact}
S{\zeta} = \frac{2i\zeta{\mathcal M}(\lambda) - |\beta|^2}{2i\zeta{\mathcal M 
(\lambda)} +
|\beta|^2} {\mathbf P}_{sym} + {\mathbf P}_{asym}.
\end{equation}
If  all  eigen-values of  the  inner  Hamiltonian are  positive (thus  embedded
  into  the  continuous  spectrum of  the  outer  operator), then the
scalar  analytic  factor  in front  of  the
${\mathbf P}_{sym}$ is an analytic function of the variable $p=\sqrt{\lambda}$
in upper half-plane $\Im \zeta > 0 $. Its  zeroes $\zeta_s$   are {\it
resonances}. For large  $\beta$,   resonances are  situated near  to the
square  roots  of the eigen-values $\lambda_s$  of the inner  Hamiltonians and
can  be  easily  calculated   in our  case via a perturbation procedure
based  on  Laurent expansion  of ${\mathcal M}$  on  powers  of  $(\lambda_s
-\lambda)$.  The  value  of  the  scattering  matrix at  the  resonance
$\zeta_0 = p_0 + i \tau_0$  can  be  calculated  as
\[
S(\zeta_0) e =  {\mathbf P}_{asym} e.
\]
One  may  assume that the  resonances  of  the  whole  quantum  system  are
known. Then  we   use  {\it resonance  amplification}, averaging
the  correlation   $\langle {\mathbf S} e, e\rangle $ of  the  scattering matrix
  with the Breit-Wigner distribution  with properly  selected parameter
$\zeta_0 = p_0 + i \tau_0$:
\[
\rho{p} = \frac{1}{\pi}\,\, \frac{1}{(p-p_0)^2 + \tau_0^2}raisebox{.5ex}{.}
\]
This
 allow us to reduce both problems to  the  calculation
of  the  averaged  correlations
via   Cauchy  integrals:
\begin{equation}
\label{BWcorr}
   \int_{R}\rho (p)\langle{\bf S}e,\, e\rangle
(p) dp = \langle{\bf S} (\zeta_0) e,\, e\rangle ,\,\,
   \int_{R}\rho (p)\langle{\bf S}^m e,\, e\rangle
(p) dp = \langle{\bf S}^m (\zeta_0)e,\, e\rangle,
\end{equation}
then averaging  the correlation  $\langle{\bf S}e,\,
e\rangle (p)$ with  a random test-vector over the above Breit-Wigner
distribution to get  the following result:
\begin{equation}
\label{BWcorr*}
   \int_{R}\rho (p)\langle{\bf S}e,\, e\rangle
(p) dp = \langle{\bf S}(\zeta_0) e,\, e\rangle  = |e_{\bot}|^2 = |{\mathbf P}_{asym} e|.
\end{equation}

Following  the probabilistic strategy in  Section~4,  we
compare  the averaged correlation $\int_{\R}\rho (p)\langle{\bf
s}e,\, e\rangle (p) dp$ of the relative  scattering matrix  in
states $S(1)$ and $S(2)$. In the first state  the relative
scattering matrix coincides with  identity,   hence averaging
gives $|e|^2$;  in the  second  state the averaging  on a  random
test-vector  gives $|e_{\bot}|^2 = |e|^2 - |{\mathbf P}_{\beta}e|^2 $.
\fi

Following  the probabilistic strategy in  Section~4,  we
compare  the  scattering matrices ${\mathbf  S}_{\beta}(\alpha_{1}^2)$ in
states $S_1$ and $S_2$. In the first state this 
matrix coincides with  identity,  hence  $$ \frac {1}{2}(I+{\mathbf  S }_{\beta}(\alpha_{1}^2))=I.$$
In the  second  state we have $$ \frac {1}{2}(I+{\mathbf  S }_{\beta}(\alpha_{1}^2))= 
\left(\begin{array}{cc}
        I & 0\\
        0 & I
         \end{array}\right) + \left(\begin{array}{cc}
        \PP_{\beta} & 0\\
        0 & \PP_{\beta}
         \end{array} \right) {\mathcal P}_{sym},$$ where  ${\mathcal P}_{sym}= \frac{1}{2} \left(\begin{array}{cc}
        I & I\\
        I & I         \end{array}\right)$  is the projection  onto the symmetric subspace of  the input space ${\mathcal E}\oplus {\mathcal E}$ consisting of vectors with equal componenets $\left(\begin{array}{c}
       e \\
        e
         \end{array}\right)$. The projections $\left(\begin{array}{cc}
        \PP_{\beta} & 0\\
        0 & \PP_{\beta}
         \end{array} \right)$ and ${\mathcal P}_{sym}$ commute and their product gives the projection onto the space spanned by $\ob = \left(\begin{array}{c}
       \beta \\
        \beta
         \end{array}\right)$. Hence,
\[\frac{1}{2} (I + {\mathbf  S }_{\beta}(\alpha_{1}^2))= \left(\begin{array}{cc}
        I & 0\\
        0 & I
         \end{array} \right) - {\mathcal P}_{\beta},\]

\noindent where ${\mathcal P}_{\beta}$ is the orthogonal projection on the subspace ${\mathcal E}$
which is collinear to the vector $\ob$ in  the input space. Therefore, in the second case, for every vector $e \in {\mathcal E}\oplus {\mathcal E}$ we have:
$$\frac {1}{2}(\langle e,\, e\rangle +\langle{\mathbf
S}_{\beta}(\alpha_{1}^2)e,\, e\rangle) =
|e|^2 - |{\mathcal P}_{\beta}e|^2.$$

   The expectation is   that if the  probability of  the
event  ${\mathcal P}_{\beta}\, e = 0$ is  zero,  then by choosing a  random
test-vector,  with  probability $1$
the  above correlation is
strictly  less  than $1$. To  obtain the  corresponding  quantitative  result we
will  assume  that  we  have  a  testing  ``device" distinguishing between
the  two  states  of  the system , which
``clicks"   if
\begin{equation}
\label{click}
  |{\mathcal P}_{\beta}e|^2 > \varepsilon|e|^2.
\end{equation}
Unfortunately,  the  above  ``device"  is not sensitive enough to
derive proper estimates for probabilities  and we need another  norm in the right-hand side of the last inequality. In our
case the input space ${\mathcal E}$ is $l_2$ with the standard orthogonal
basis
 ${\mathbf x} \longrightarrow \left\{ x_m \right\}_{m=0}^{\infty}.$ Following (\cite{Coins})  we consider  the
discrete  Sobolev classes and norms introduced in Section~4,
(\ref{norm1}) and (\ref{norm2}) in  order  to  define the case
when the  ``device" clicks.  Next we assume
      that  the (complex) increments  $x_m - x_{m-1}$ are
      independent.
      We are going to use, together with $l_2$
      two more  spaces of  test-vectors. Both  are
      stochastic spaces of  all  trajectories  ${\bf x} (t)$ of a
      Brownian particle  on  the  complex  plane along different
      discrete  sequences of intermediate   moments of
      time (``stops"): the
equidistant sequence
       $t_l = \sum_{m=1}^l \delta_m$,  for  the  first  space,
      and the perturbed    sequence
      $\tilde{t}_l = \sum_{m=1}^l {\Delta}_m$, for  the  second  space.
       Both  spaces  are  equipped  with proper Wiener
      measures  $W,\,\tilde{W}$  (see \cite{stroock}).  The measure $W$ is
      defined  on the  algebra  of  all finite-dimensional cylindrical sets
      $C_{_{_{\Delta_1,\Delta_2,\dots ,\Delta_{N}}}}^{t_1,t_2,\dots
      ,t_{N}}$ of  trajectories with fixed  initial  point
      $x_0= 0$ and   ``gates" $\Delta_l,\, l = 1, \ldots, N$ (which are
       open  discs  in  the  complex  plane):
     \[
      C_{_{_{\Delta_1,\Delta_2,\dots ,\Delta_{N}}}}^{^{^{t_1,t_2,\dots
      ,t_{N}}}} = \left\{ \x \mid x_{t_l}\in \Delta_l, l=1,2,\dots ,N
      \right\},
      \]
      \noindent via multiple convolutions  of the Green  functions
      $G(x_{l+1},t_{l+1}\big| x_l,t_l) $  corresponding
      to  the sequence  $\delta_{l+1} = t_{l+1}- t_l$:

        \[
      W (C_{_{_{\Delta_1,\Delta_2,\dots
      ,\Delta_{N}}}}^{^{^{t_1,t_2,\dots ,t_{N}}}}) =
      \phantom{xxxxxxxxxxxxxxxxxxxxxxxxxxxxxxxxxxx}
      \]  \\[-10ex]

      \begin{equation}
      \displaystyle
        \label{cylinder} \frac
        { \int\dots\int_{\Delta_{N},\Delta_{N-1},\dots
      ,\Delta_{1}} \frac{dx_1 dx_2 \dots dx_{N}}{\pi^{\frac{N}{2}
      \sqrt{\delta_{N} \delta_{N-1}\dots \delta_1 }}}
      \, e^{^{-\frac{|x_N - x_{N-1} |^2}{\delta_N}}}\dots e^{^{-\frac{|x_1
      - x_{0} |^2}{\delta_1}}} }{ \int\dots\int_{\R_{N}, \R_{N-1},\dots
      ,\R_{1}} \frac{dx_1 dx_2 \dots dx_{N}}{\pi^{\frac{N}{2} \sqrt{
      \delta_{N} \delta_{N-1}\dots \delta_1 }}} \, e^{^{-\frac{|x_N -
      x_{N-1} |^2}{\delta_N}}}\dots e^{^{-\frac{|x_1 - x_{0}
      |^2}{\delta_1}}} },
      \end{equation}

      \noindent where $\R_{N} = \R_{N-1} = \dots
      = \R_{1} = \R$. Using the convolution formula,
      the
      denominator of   (\ref{cylinder})  can  be  reduced to the  Green
function
       $ G(x_N,t_N \mid 0, 0)$,
      for  any  $\tau \in (s,t)$:
      \[  G(x,t \mid y,s) = \int_{-\infty}^{\infty}  G(x,t \mid \xi,\tau)
        G(\xi,\tau \mid y,s) d\xi.
      \]
In a similar way we can define    the  Wiener  measure for  trajectories
corresponding to
   the ``perturbed" sequence   $\tilde{t}_l$.

       \par
   In what follows we are  going to use the  {\it absolute continuity}
    of  the {\it perturbed} Wiener  measure  $\tilde{W}$ with respect
     to the  non-perturbed one $W$: for
      every
      $W$--measurable set
      $\Omega$,

      \begin{equation}
      \label{relW}
      \tilde{W}(\Omega) =
      \frac{1}{\prod_{l=1}^{\infty}\sqrt{\Delta_l}}
      \int_{\Omega} e^{- \sum_{m=1}^{\infty}
      \frac{1-{\Delta}_m}{{\Delta}_m} \mid x_m -
      x_{m-1}
      \mid^2 } dW.
      \end{equation}

      Further we consider  the class  of
       {\it quasi-loops}, that  is the class of  all  trajectories  of  the
      ``perturbed  process"   which begin  from
      $ (x_0, \tilde{t}_0) = (0,0 )$  and  for  any  $t$,
       $\max_{0< s <t} |x_s|^2 < Ct$. We note that
       \begin{itemize}
      \item every $\x \in l_2^1$ is a quasi-loop (with $C = \,  \mid \x
      \mid_1^2$),
      \item due to the reflection principle {\rm (see \cite{stroock}, p.
221)},
      the class of
      all quasi-loops has Wiener measure one, both in respect of
$W,\,\,\tilde{W}$.
      \end{itemize}
\if01
We  assume  that  the  ``device'' is  applied to  the correlations
averaged  over the  Breit-Wigner distribution with properly chosen
parameters ${\mathbf S}(\zeta_0)\beta = 0$. 
Then
\[
   \int_{{\mathbf R}}\rho (p)\left[\langle {\bf x},\, {\bf x}\rangle - \langle{\bf
S}(e){\bf x},\,
   {\bf x}\rangle \right] (p) dp = \langle {\bf x},\, {\bf x}\rangle -
   \langle{\bf S}(\zeta_0) {\bf x},\, {\bf x}\rangle
    =1 - |{\bf x}_{\bot}|^2 = |{\mathbf P}_{\beta} {\bf x}|^2.
\]
\fi

We  assume  that  the  device  clics, if the  result of  averaging
exceeds a certain level defined  by  the  above  norm $\parallel \cdot
\parallel^2_1$:
\[
|{\mathcal P}_{\beta} {\bf x}|^2 > \, \,\, \varepsilon
      \parallel\x \parallel^2_1.
\]
This  device  cannot identify the
        state  of  the  system  from  the  observation of  the
      Breit-Wigner  averaged  correlation between the  input  and  output
      of  a  single  act  of  scattering when presented  a randomly  chosen
      input ${\bf x}\in {\mathcal  E}$  if
     $
     |{\mathcal P}_{\beta} {\bf x}|^2 < \varepsilon \parallel{\bf x}\parallel^2_{1}.
      $
      This means that the test-vector $\x$ belongs
       to  the {\it indistinguishable set}
      \begin{eqnarray}
       \label{sensor}
      \IF_{\varepsilon} &  = & \left\{ \x \in l_2 \cap l_2^1 \,,\,
|{\mathcal P}_{\beta} {\bf x}|^2
      < \, \,\, \varepsilon\,\, \left(  \sum_{_{_{m =
      1}}}^{\infty}
      \frac{1-{\Delta}_m}{{\Delta}_m}\, |x_{_m} -
      x_{_{m-1}}|^2\right)\right\}\\ [2ex]\nonumber
      &  = & \{ \x \in l_2 \cap l_2^1 \,,\, |{\mathcal P}_{\beta} {\bf x}|^2
      < \,
     \,\, \varepsilon
      \parallel\x \parallel^2_1 \}.
      \end{eqnarray}
    Though technically we may easily consider, with Breit-Wiegner
    averaging, the iterated scattering processes described by the powers
    ${\bf S}^m$ of  the  scattering matrix, we  will analyze now
    the independent  single  acts of  scattering. In this  case
    the  indistinguishable  set  depends only upon  the  positive  number
    $\varepsilon$, the  vector $\ob \in {\mathcal  E}\oplus {\mathcal  E}$    defining  the
      interaction  in  the  quantum  system,  and  the  sequence
      ${\Delta}$. Without  loss of  generality  we  may
      assume that  the  vector  $\ob$
      has  all  non-zero  components $\ob_l \neq 0$. We assume  that
      the vector ${\bf b} = \left\{ b_l \right\}_{l=1}^{\infty},\,\,
       b_l = \sum_{m=l}^{\infty}\ob_m$
        belongs to $l_2$:
      \begin{equation}
      \label{betacond}
      |b|^{2}_{l} = \sum_{m=1}^\infty  m^2 |\ob_m|^2 < \infty.
      \end{equation}

    Our main result reads:  {\it If  the condition (\ref{betacond})
     is  satisfied, then the  Wiener  probability
     $\tilde{W} ( \IF_{\varepsilon,1})$ of  the  indistinguishable
     set  $ \IF_{\varepsilon, 1}$    corresponding to  a  single  act
     of scattering  is finite  and  is  estimated as }
     \begin{equation}
     \label{inssisest}
     \tilde{W} ( \IF_{\varepsilon}) <
     \frac{\sqrt{\varepsilon}\,|\ob|}{\prod_{l=1}^{\infty} \sqrt{\Delta_l}
     \sqrt{\varepsilon |\ob|^2+ |b|^2}}\raisebox{.5ex}{.}
\end{equation}

       Following the  calculation presented  in  \cite{Coins}, we
        approximate  the  indistinguishable  set  with
        finite-dimensional  cylinder sets  and  reduce  the
       estimation of    $\tilde{W} ( \IF_{\varepsilon,1}) $      to  the
       calculation of  a  Wiener  integral  with respect to  the   $W$
measure
       on  trajectories  associated  with  ``equidistant stops''.
       We  have:

      \[\tilde{W} ({\IF}_{\varepsilon, 1 })
      \phantom{xxxxxxxxxxxxxxxxxxxxxxxxxxxxxxxxxxxxxxxxxxxxxxxxxxxxxxxxx}
      \]\\[-7ex]
      \begin{eqnarray*}
      & \le & \,\,
      \int_{\mbox{quasi-loops}, x_0 = 0\,
      |\langle {\bf x},\ob\rangle|^2 < \varepsilon\parallel {\bf x}
\parallel^{2}_{1}} \,\,
      d{\tilde{W}} \\
      &  =  & \lim_{C\to\infty} \int_{sup_{s\leq l}|x_s| <
C\sqrt{l}, l\leq N, \, x_0 = 0
      |\langle {\bf x},\ob\rangle|^2 < \varepsilon\parallel {\bf x}
\parallel^{2}_{1} }\,\,
      d{\tilde{W}}  \\
      & =  & \frac{1}{\prod_{l=1}^{\infty}\sqrt{ \Delta_l}}\,\,
      \lim_{C\to
      \infty}\lim_{N \to \infty}
      \\
      &&
      \frac{\int_{sup_{s\leq l}|x_s| < C\sqrt{l}, l\leq N, \, x_0 = 0}
      \int \int \dots \int_{-\infty}^{\infty} dx_1
      dx_2  \dots  dx_N
       e^{-\frac{|x_N - x_{N-1}|^2 }{\Delta_{N}}}
        \dots
         e^{-\frac{|x_1 - x_{0}|^2 }{\Delta_{1}}} }
      {\pi^{N} \int_{_{|x_N| < C\sqrt{N}}} d x_N e^{-\frac{x_N -
x_0}{\tilde{t}_N}} } \\
      \end{eqnarray*}
      \begin{eqnarray*}
      &&\leq
       \frac{1}{\prod_{l=1}^{\infty} \sqrt{\Delta_l}}\,\,
      \lim_{C\to
      \infty}\lim_{N \to \infty}
      \\
      &&
      \frac{\int_{sup_{s\leq l}|x_s| < C\sqrt{l}, l\leq N, \, x_0 = 0}
      \int \int \dots \int_{-\infty}^{\infty} dx_1
      dx_2  \dots  dx_N
       e^{-\sum_{s=1}^N\frac{|x_s - x_{s-1}|^2 }{\delta_s} }
       e^{{\mathcal  B}(x_0,x_1,x_2\dots x_N)}}
      {\pi^{N}  \int_{|x_N| < C\sqrt{N}} d x_N
      e^{-\frac{|x_N - x_0|^2}{\tilde{t}_N}} } \raisebox{.5ex}{.}
      \end{eqnarray*}

      The integrand  of the inner integral in the numerator  contains
      the exponential  factor
      \[
        e^{\mathcal  B}=
        e^{ -(1-\Delta_{N})\frac{|x_N - x_{N-1}|^2 }{\Delta_{N}}
        - (1 - \Delta_{N-1}) \frac{|x_{N-1} - x_{N-2}|^2 }{\Delta_{N-1}}
        - (1 - {\Delta_1} )\frac{|x_1 - x_{0}|^2}{\Delta_1}},
      \]
   which  can be  estimated  due to (\ref{sensor}) by the
   exponential:
   $$e^{-\frac{1}{\varepsilon} |{\mathcal P}_{\ob} {\bf x}|^2} =
    e^{-\frac{1}{\varepsilon |\ob|^2 } |\langle{\bf
x},\,\ob\rangle|^2}.$$
    Using this equality, the exponential  in the  numerator
    can  be  estimated from below  by  the quadratic  form
    \begin{equation}
    \label{expo}
    -\sum_{m=1}^N |x_m - x_{m-1}|^2 - \frac{1}{\varepsilon |\ob|^2}
     |\langle{\bf x},\,\ob\rangle|^2.
    \end{equation}
This  quadratic  form  can be simplified  using new vector
variables  $\xi_m = x_m - x_{m-1}$:
\[
\langle{\ob,\,\bf x}\rangle = \sum_{m=1}^{\infty} x_m
\bar{\ob}_m = \sum_{m=1}^{\infty} \xi_m
\overline{\sum_{l=m}^{\infty}{\ob_l}}.
\]
Recall that  the  vector $ {\bf b}, \, b_m = \sum_{l=m}^{\infty}
{\beta}_l\,\,$ belongs to $l_2$. Then  the  quadratic  form in the
exponent  of  the numerator  can  be  presented  as  a  quadratic
form of an operator
\[\langle \xi, A_{\varepsilon} \xi \rangle =
|\bf{\xi}|^2  + \frac{1}{|\ob|^2 \varepsilon} |\langle b,{\bf
\xi}\rangle| = \langle \xi, \left(I + \frac{|b|^2}{|\ob|^2
\varepsilon} {\mathbf P}_b \right)\xi \rangle ,
\]
where ${\mathbf P}_b$ is  the  orthogonal projection onto the
one-dimensional  subspace in $l_2$ spanned by  the  vector ${\bf b}$. The
  ratio of  the $N$-dimensional Gaussian  integral in the  numerator,
normalized by the factor $\pi^{-N/2}$ and  the Gaussian integral
in the  denominator can be expressed as
\[
\frac{1}{\pi^{N/2}}\int \int  \dots \int e^{-\langle \xi,
A_{\varepsilon} \xi \rangle} d\xi_1 d\xi_2 \dots d\xi_N =
\frac{1}{\sqrt{\mbox{det} A_{\varepsilon}}} =
\frac{\sqrt{\varepsilon}\|\ob|}{\sqrt{\varepsilon|\ob|^2 +
|b|^2}}\raisebox{.5ex}{.}
\]
  Finally,   we  obtain the  announced  result by taking into
  account  the omitted  factor $\prod_l
\sqrt{\Delta_l}$.

\section*{Acknowledgement} We thank Radu Ionicioiu for his comments
on a draft form of this paper. Calude and Pavlov have been supported in part
by the The Vice-Chancellor's University Development Fund 23124/2002.

\end{document}